# INSTRUCTOR PERSPECTIVES OF MOBILE LEARNING PLATFORM: AN EMPIRICAL STUDY

Muasaad Alrasheedi, Luiz Fernando Capretz and Arif Raza

Department of Electrical and Computer Engineering, Western University, London, Ontario, Canada

## ABSTRACT

*Mobile learning (m-Learning) is the cutting-edge learning platform to really gain traction, driven mostly by the huge uptake in smartphones and their ever-increasing uses within the educational society. Education has long benefitted from the proliferation of technology; however, m-Learning adoption has not proceeded at the pace one might expect. There is a disconnect between the rate of adoption of the underlying platform (smartphones) and the use of that technology within learning. The reasons behind this have been the subject of several research studies. However, previous studies have mostly focused on investigating the critical success factors (CSFs) from the student perspectives. In this research, we have carried out an extensive study of the six factors that impact the success of m-Learning from instructors' perspectives. The results of the research showed that three factors – technical competence of instructors, Instructors' autonomy, and blended learning – are the most important elements that contribute to m-Learning adoption from instructors' perspectives.*

## 1. INTRODUCTION

The mobile phone has perhaps become the most widely adopted technology of the modern world, reaching new heights of use and technological advancement. The almost continual development of the platform, including its features and capabilities, means that for a learning platform based upon these phones, as m-Learning is, there is an expectation of rapid adoption rates throughout the world. The education sector is one that has been proactive in its use of new technology, both in terms of learning and teaching, and the use of mobile technology and the m-Learning platform is definitely on the rise.

It is clear that m-Learning offers something new, beyond that which is offered by e-learning, notably the mobility aspect of the platform. This provides a flexibility for learners – in terms of location and also of time, pace, and space – that is not readily achieved with any other non-mobile device or platform [1]. Additionally, m-Learning offers a unique opportunity for collaborative learning, with the mobile phone platform offering a much wider level of interaction with other students and teachers, even when they are not in a formal classroom situation or even in the same location. This combination of mobility and collaborative learning is what makes the m-Learning platform so unique and makes it stand out from existing learning methods, whether the traditional face-to-face type method or other technology-based platforms such as e-learning [2].

It is the diversion between the take up of the mobile platform itself and the take up of m-Learning that is the most interesting thing about the platform, for the wrong reasons obviously, even among demographics with very high levels of technology acceptance. A large proportion of the research available cites high interest in exploring m-Learning[3]; however, adoption within the educational institutions has nevertheless remained slow [4]. The higher education sector, in fact, shows one of the slowest adoption rates of all, despite students repeatedly showing they are receptive to using mobile devices as a tool during learning, devices they are extremely comfortable with [5].

Researchers have offered reasons for this, centering on resistance from the educators themselves, who are frequently uncomfortable with the technology or simply lacking in the knowledge needed to use the platform with confidence. This is exacerbated by the lack of any adoption framework, and the continual changing of the platform technology itself, the smartphones, as well as ongoing concerns regarding privacy and security, which have all contributed to the slow rates of adoption[6]. It is, therefore, clear that instructors attitudes towards the idea of m-Learning are a major barrier to overcome, and that is why the opinions of educators is the subject of this study. With the future of educational technology heading towards ubiquitous computing, the ability of educators to maximize the use technology to enrich both teaching and learning within the classroom is essential. To this end, having teacher training focus on the mobile vision and competencies will be crucial to the transformation of the pedagogy. Additionally, the issues and apparent problems facing the technology need to be addressed prior to the introduction of large-scale projects[7].

It is a characteristic of modern mobile technology to have small sizes, relatively cheap costs, and ubiquity in combination with the proliferation of accessible wireless networks; this has focused researchers towards the possibilities that the combination of this technology with applications represents (Peng et al., 2009)[7]. Although the idea of having support for high school teachers from mobile technology has been investigated, including the development of an interface design, a complete system with the requisite support system necessary for successful implementation is not available to high school teachers at present. To address this, the study looks towards a Ubiquitous Performance Support System for Teachers (UPSST), with the intention of providing a plan from design to implementation that is focused on improving the performance of high school teachers. Two high-school homeroom teachers were the initial users of the UPSST, and another 12 homeroom teachers participated in testing the development [8].

In the next section is presented the literature review where several relevant aspects related to mobile learning and perception have been discussed. Section 3 presents the research model and the hypotheses that would be tested. Section 4 presents the research methodology. Section 5 presents data collection and the measuring instrument. Then Section 6 presents analysis of hypothesis tests and results. Section 7 presents a discussion of the results then the limitations of the present study in section 8. Section 9 presents the conclusion.

## 2. LITERATURE REVIEW

Educational institutions focus a significant proportion of their time and effort on ensuring that students are learning; this is the driving force behind the face-to-face interactions during teaching and examinations. This need to ensure that students are learning is improved by wireless technology, where instructors are expected to impart educational learning to the same standards with the same caliber of student if m-Learning is going to be considered a mainstream educational platform [9].

Looking at student interaction with the platform, the work of Brett [10] noted that the user experience of the platform remained very positive where such use aligned with student need, and with students owning ever more sophisticated technology themselves. With educational institutions facing ever tighter fiscal limitations, the shift to student-supplied devices would seem beneficial.

However, for the m-Learning system to succeed various CSFs should be considered. This study offers a systematic approach to analyzing instructors' perceptions of a successful m-Learning platform that can be emulated in other studies to understand the CSFs of m-Learning. We studied and empirically analyzed the impact of six CSFs that have had the most effect on instructors' perceptions based on our previous research[11]; those factors are the technical competence of instructors, instructors' autonomy, user-friendly application design, assimilation with curriculum, instructor community development, and blended learning.

Considering the first factor, the technical competence of the instructors, Volery and Lord [12] report that the instructors must have adequate skills with the technology that will enable them to carry out teaching through the Internet. A lack of technological skills by instructors will be a significant hindrance to the adoption of the new learning technology [13].

Another important factor for successful adoption of the system of m-Learning is instructors' autonomy. One of the critical factors for success in adoption of an m-Learning system is the way the instructor uses m-Learning. When instructors who decide to use the new system of m-Learning encourage students to start appreciating the value in adopting the system, the influence of the instructor will motivate students to use the technology in their studies. This happens because social influence is a determinant for adopting the new technology, and it is important for students to receive encouragement to adopt new technology within their learning setting [14].
The assimilation of the m-Learning system with the curriculum is also a key factor in the success of adopting the m-Learning system. If the ministry of education permits the m-Learning system to be included in the education system the adoption of m-Learning will be quite successful. However, if the educational ministry does not acknowledge the system, it will be very challenging to adopt the mobile system in education [15].

In addition, the availability of a user-friendly application design is another major factor that influences the adoption of the m-Learning system. Unlike standard computers, the user interfaces of mobile devices are extremely varied and designing a common user interface present a challenge [16]. This design should be able to be used with ease by both students and instructors when dealing with the system of m-Learning. Additionally, user-friendly design is perceived to positively correlate with the perceptions of instructors. That is, for users to choose the platform of m-Leaning, a user-friendly design is essential [17].

Instructor community development, that is, using the platform of m-Learning to connect with instructors and other learners [18], also plays an important role in adopting the m-Learning system. With this connection, adoption of m-Learning becomes effective because the instructor wants to keep contact with the students and that is only possible through the use of m-Learning [19]. Blended learning is another CSF in the adoption of the new technology. Al-Busaidi and Al-Shihi [20] argue that students are involved in blended learning which they learn at home through the home-grown LMS. Blended learning is a key factor in adopting the system of m-Learning since students using blending learning cannot afford to do it without the use of the new learning technology of m-Learning.

## 3. RESEARCH MODEL & HYPOTHESIS

The This research is intended to present a research model for the assessment and analysis of the six factors (CSFs) that affect instructors' perspectives regarding m-Learning adoption within higher education.

Figure 1 below shows the research model diagram. The model derives its theoretical foundations by combining the previous work by [11][16]. The model uses six CSFs: 1. The technical competence of instructors, 2. Instructors' autonomy, 3. User-friendly application design, 4.

Assimilation with curriculum, 5. Instructor community development, and 6. Blended learning. The dependent variable of this study is m-Learning adoption according to instructors' perceptions of m-Learning. The six independent variables are referred to as CSFs hereafter.

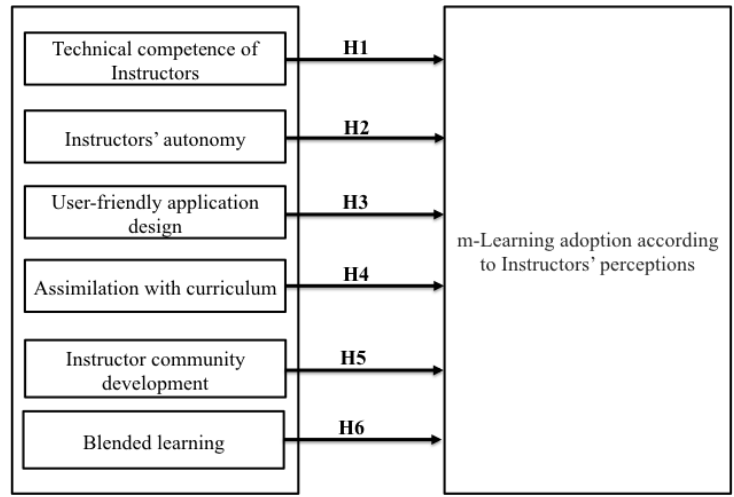

Figure 1. Research model – critical success factors affecting the success of m-Learning from instructors' perspectives.

Since instructors are at the very core of the learning process, it is essential that their views and ideas regarding a new learning platform are fully understood. In our previous work [11] we found six factors that affect the overall attitude towards m-Learning. A detailed survey has been constructed to enable us to determine the CSFs for m-Learning from the instructor perspectives, with the final objective of this research aiming to offer a response to the following question:

To what extent do various CSFs impact m-learning adoption from instructor perceptions?

The multiple linear regression equation of the model of the answer is represented as follows:

m-Learning adoption from instructors' perceptions = $C_0 + C_1f_1 + C_2f_2 + C_3f_3 + C_4f_4 + C_5f_5 + C_6f_6$. - Equation (1)

In the equation $C_0$, $C_1$, $C_2$, $C_3$, $C_4$, $C_5$, and $C_6$ are coefficients and $f_1$, $f_2$, $f_3$, $f_4$, $f_5$, and $f_6$ are the six independent variables.

To empirically investigate the research question, the six hypotheses are presented below with a belief that they all positively affect m-Learning adoption according to the instructors' perceptions as presented:

Hypothesis 1. Technical competence of instructors positively affects the m-Learning adoption according to instructor perceptions.
Hypothesis 2. The extent of instructors' autonomy has a positive relationship with the m-Learning adoption according to instructors' perceptions.
Hypothesis 3. User friendly design of the m-Learning platform is positively related with the m-Learning adoption according to instructors' perceptions.
Hypothesis 4. Assimilation with the curriculum will directly affect the m-Learning adoption according to instructors' perceptions.
Hypothesis 5. Perception of increased opportunities for learner community development and the m-Learning adoption according to instructors' perceptions are positively related.

Hypothesis 6. The blended learning possibility will positively affect the m-Learning adoption according to instructors' perceptions.

# 4. RESEARCH METHODOLOGY

The instructors are a vital component of the learning platform, not only as one of the two primary user groups of the platform, but also as the mentors for the other primary user group, the actual learners. In addition, being the designers of the course and the disseminators of the material makes instructors the most important stakeholder in the m-Learning adoption process. The present study looks to collate and analyze the views of instructors in a systematic method.

Our methodology is presented in Figure 2 below. First, we systematically identified the factors contributing directly or indirectly to m-Learning adoption from the instructors' perspectives. In order to do an empirical investigation of the key factors from the instructors' perspectives, a research model was developed based on the key factors shown in Figure 1. Then a questionnaire was prepared and we conducted a survey to assess each key factor. Finally, we performed a statistical analysis of data on instructors' perspectives. The data analysis was performed using Minitab v.17 as our quantitative analysis tools.

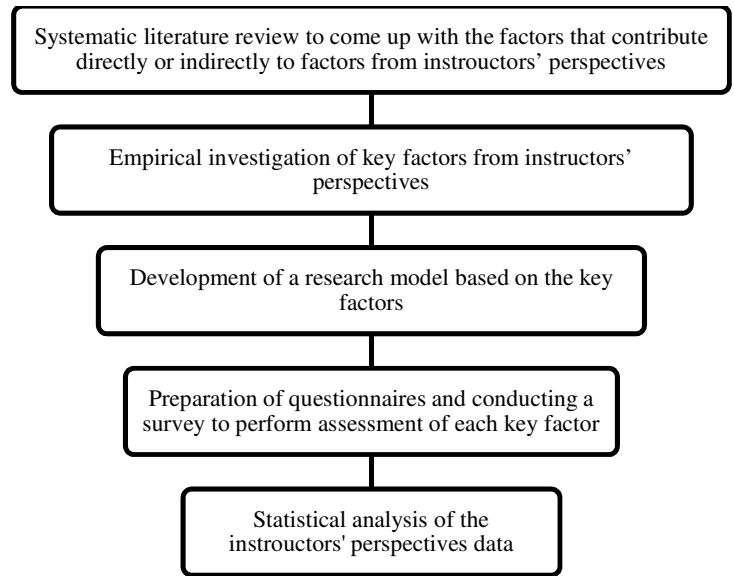

Figure 2. Steps representing the research methodology.

# 5. DATA COLLECTION AND THE MEASURING INSTRUMENT

To collect the data, we used an online survey tool (SoGoSurvey). The questionnaire was sent to various instructors teaching different undergraduate and post-graduate courses in five universities in Saudi Arabia. We received a total of 64 completed questionnaires.

The measuring instruments presented in (Appendix I) were used to study the perceived level of instructors' satisfaction as well as the extent to which these CSFs were important for the instructors in adopting m-Learning. The questionnaire required participants to indicate the extent of their agreement or disagreement with statements using a five-point Likert Scale. For all of the items associated with each variable, the scale ranged from (1= Strongly Disagree, 2= Disagree, 3= Neither Agree or Disagree, 4=Agree, 5= Strongly Agree).

Our questionnaire had three parts:

1. The first part was used to determine the general profile of the respondents and consisted of questions regarding their gender, age group, and the level of students that they teach.
2. The second part was used to determine the extent to which instructors have accesses to mobile devices and the Internet, and their experience in using these devices in teaching.
3. The third part was used to determine the different factors that affect user perception of the m-Learning platform as below: Question 1- (technical competence of instructors), Question 2- (Instructors' Autonomy), Question 3- (User Friendly design), Question 4- (Assimilation with curriculum), Questions 5-8- (Learner community development), Question 9 (Blended learning), and Questions 10-12 (Cumulative overall instructors' perspectives).method.

## 5.1. Data analysis procedure

Firstly we started our data analysis by a descriptive analysis of demographic distribution of the population. Next, in order to analyze the research model and test the hypotheses, the data analysis procedure involved three phases. In phase one, a parametric correlation was found between the dependent and independent variables to see if any of the variables, i.e., hypotheses, can be accepted or rejected. The second phase was conducted to compute a non-parametric correlation using the same set of data in order to reduce the threat to external validity) [21]. Finally, the third phase involved testing the hypotheses by using the Partial Least Square (PLS) technique.

## 5.2. Demographic distribution of the population

Here we look at the distribution of demographics within the population. As mentioned, the total population comprised 64 instructors. Of this, 47 were male and 17 were female. Further, as mentioned earlier, the population comprised instructors from different universities. The distribution was reasonably uniform. Only one of the instructors was under 25 years of age. A majority, i.e., 36 of the instructors, were between 36-55 years of age. The next largest age group was 26-35 years, which was 21 of the instructors. Only 6 instructors were over 55 years of age. An overwhelming majority of the instructors, i.e., 61 out of 64, were employed full-time; the remaining were employed part-time. In terms of the teaching levels, 48 instructors or 75% of the research population taught undergraduate classes, while the remaining 16 instructors or 25% of the research population taught post-graduate classes.

An essential component of the analysis of the demographics was to establish the level of mobile phone use within the user group, and the survey provided interesting data in this regard. All instructors owned a mobile phone, and a majority owned several devices, i.e., 59 of the instructors owned a smartphone or a personal digital assistant (PDA). Additionally, 55 instructors owned a desktop PC, while a significant majority, 62 instructors, owed a laptop, tablet, or notebook. All instructors had Internet installed on at least one of these devices, and a significant majority, i.e., 59, of the instructors had the Internet installed on their mobile phones.

The extent of adoption of both mobile phones and Internet access among the instructors was incredibly high, displaying both awareness of the tools available and wide adoption of the mobile phone as a tool for accessing the Internet.

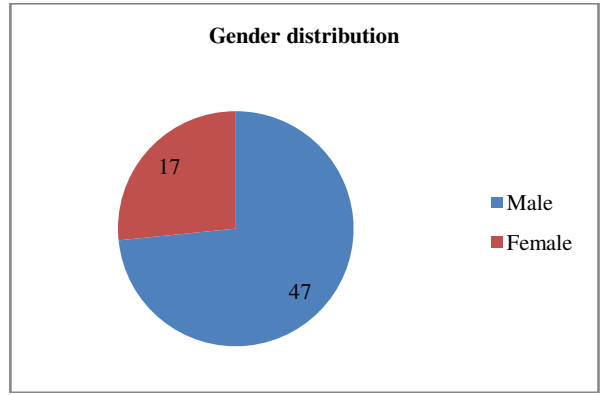

Figure 3: Respondents' gender distribution

## 5.3. Reliability and validity analysis of measuring instrument

This m-Learning survey was created using a series of questions that looked to evaluate the attitude of the instructors towards the adoption of m-Learning. Five of these questions were straightforward involving single-item measurements. However, two of the questions involved multi-item rating scales: Instructors community development and the overall instructors' perceptions; these two questions were also measured using three-item measurement. In all of these cases, it is important to assess the reliability of the measurement scales. This is done to quantify the reproducibility of a measurement and is performed using an internal consistency analysis – calculating the Cronbach's alpha. The limit of the satisfactory levels for this reliability coefficient has been determined by various types of research. Most of the existing work cites the Ven and Ferry study, which considers the coefficient of 0.55 and higher as satisfactory [22]. Other work, such as the study by Osterhof [23], however, suggest that a reliability coefficient of 0.6 minimum satisfaction is more appropriate. For this study, the reliability coefficient in all cases is > 0.7, to offer a reliable measuring instrument. Table 1 illustrates the values of Cronbach's alpha and Principal Component Analysis (PCA) Eigen values applicable to the factors in question.

Table 1.  Cronbach's alpha for multi-measuring rating scales.

| Success Factors | Item Numbers | Cronbach's alpha | PCA Eigen Values |
|---|---|---|---|
| Technical competence of instructors | 1 | 0.8145 | 1.688 |
| Instructors Autonomy | 2 | 0.7936 | 1.658 |
| User friendly design | 3 | 0.7569 | 1.609 |
| Assimilation with curriculum | 4 | 0.7471 | 1.601 |
| Instructors' community development | 5-8 | 0.7574 | 1.614 |
| Blended learning | 9 | 0.8218 | 1.700 |

Validity is defined as the strength of interference between the value of a measurement and its true value. Comrey and Lee's [24] PCA was performed for all six CSFs, and reported in Table 1. Eigen value [25] has been used as a reference point, to observe the construct validity, using PCA. We used the Eigen Value One criterion, which is known as the Kaiser Criterion [26]; [27] that indicated any component having an Eigen value more than one would be retained.  The results of the study show that the Eigen-value analysis of all 6 variables form a single factor, as seen in Table 1. Statistical analysis, therefore, shows that the convergent validity of the instrument for instructors' perspectives on m-Learning adoption can be considered sufficient.

## 6. HYPOTHESIS TESTS AND RESULTS

The significant hypotheses, H1-H7, were analyzed within the research model using three statistical methods within three distinct phases. Phase I consisted of normal distribution tests and parametric statistics, while phase II used non-parametric statistics. Both parametric and non-parametric statistical approaches were used to reduce the threats to external validity. As our measuring instrument had multiple items for all the six independent variables as well as the dependent variable (shown in Appendix I), the ratings by the respondents were added up to get a composite value for each rating. Tests were conducted for hypotheses H1-H6 using parametric statistics by determining the Pearson correlation coefficient. For non-parametric statistics, tests were conducted for hypotheses H1-H6 by determining the Spearman correlation coefficient. To increase the reliability of the results, hypotheses H1-H6 of the research model were tested using the PLS technique in Phase III. The results of the statistical calculation for the Pearson correlation coefficient are shown in Table 2 below. It is established that lower p-values signify a higher chance of rejecting the null hypothesis, and, therefore provide results of more meaningful statistical significance [28]. Here all p-values were below 0.05, demonstrating that the results hold significance.

Table 2.  Hypothesis testing using parametric test.

| Hypothesis | Critical Success Factor | Pearson Correlation Coefficient | Spearman Correlation Coefficient |
|---|---|---|---|
| H1 | Technical competence of instructors | 0.689* | 0.592* |
| H2 | Instructors' Autonomy | 0.658* | 0.627* |
| H3 | User friendly design | 0.610* | 0.582* |
| H4 | Assimilation with curriculum | 0.601* | 0.564* |
| H5 | Instructors' community development | 0.615* | 0.552* |
| H6 | Blended learning | 0.701* | 0.650* |

* Significant at P < 0.05.

The results of the statistical calculation for the Pearson correlation coefficient are shown in Table 2 below. It is a commonly known fact that the lower the p-value the better chance there is of rejecting the null hypothesis and, hence, the more significant is the result in terms of its statistical significance [28]. In the present case, all the p-values in both (Pearson Correlation Coefficient and Spearman Correlation Coefficient) are < 0.05. This indicates that the results are significant.

The Pearson correlation coefficient between the technical competence of instructors towards m-Learning adoption was positive: 0.689 at P < 0.05, and, hence, hypothesis H1 is justified. For H2, the relationship between instructors' autonomy and m-Learning adoption, the Pearson correlation coefficient, was 0.658 at P < 0.05, and, hence, it was found to be significant as well. Furthermore, hypothesis H3 was accepted based on the Pearson correlation coefficient of 0.610 at P < 0.05, which represents the relationship between user-friendly design and m-Learning adoption. Similarly, hypothesis H4, the relationship between assimilation with curriculum and m-Learning adoption, the Pearson correlation coefficient, was 0.601 at P < 0.05; hence, it was found to be significant and was accepted as well. Likewise, hypothesis H5 was accepted based on the Pearson correlation coefficient of 0.615 at P < 0.05, which represents the relationship between instructors' community development and instructors' perceptions towards m-Learning. Finally, the Pearson correlation coefficient between blended learning and m-Learning adoption was positive 0.701 at P < 0.05, and, thus, hypothesis H6 is accepted. Hence, as observed and reported, all hypotheses – H1, H2, H3, H4, H5, and H6 – were found to be statistically significant and were accepted.

## 6.1. Phase II

In the second step, non-parametric statistical testing was performed by examining the Spearman correlation coefficient including the individual independent variables, all CSFs, and the dependent variable, m-Learning adoption according to instructor's perceptions, as shown in Table 2.

In phase II, a non-parametric statistical testing was conducted by examining the Spearman correlation coefficients between individual independent variables (CSFs) and the dependent variable (m-Learning adoption). The results of the statistical calculations for the Spearman correlation coefficients are also displayed in Table 2. The Spearman correlation coefficient between the technical competence of instructors and m-Learning adoption according to instructors' perceptions was positive 0.592 at P < 0.05, and, hence, hypothesis H1 was justified. For hypothesis H2, which examined the relationship between instructors' autonomy and m-Learning adoption, the Spearman correlation coefficient of 0.627 was observed at P < 0.05, which indicates that this hypothesis was significant. Moreover, hypothesis H3 was accepted based on the Spearman correlation coefficient of 0.582 at P < 0.05, representative of a statistically significant relationship between user-friendly design and m-Learning adoption. For hypothesis H4, which involves the relationship between assimilation with curriculum and m-Learning adoption, the Spearman correlation coefficient was 0.564 at P < 0.05, which means it was found to be significant; consequently, it was accepted, also. Similarly, hypothesis H5 was accepted based on the Spearman correlation coefficient of 0.552 at P < 0.05, which represents the relationship between instructors' community development and m-Learning adoption according to instructors' perceptions. The last hypothesis, H6, was accepted also, based on the Spearman correlation coefficient of 0.650 at P < 0.05, which represents the relationship between Blended learning and m-Learning adoption according to instructors' perceptions. Consequently, as observed and reported, all hypotheses H1, H2, H3, H4, H5, and H6 were found to be statistically significant and were accepted.

## 6.2. Phase III

For cross validation of the results that were obtained during Phases I and II, the PLS technique was utilized during Phase III. As put forward by Fornell and Bookstein [29], the PLS technique is incredibly useful in a variety of situations including complexity, non-normal distribution, low theoretical information, and small sample sizes, and the adoption here ensures the increased reliability of the results.

Table 3. Hypotheses testing using Partial Least Square regression (instructors' Perspectives)

| Hypothesis | Critical Success Factor | Path Coefficient | R2 | F-Ratio |
|---|---|---|---|---|
| H1 | Technical competence of instructors | 0.64 | 0.474 | 55.89* |
| H2 | Instructors Autonomy | 0.64 | 0.433 | 47.35* |
| H3 | User friendly design | 0.58 | 0.371 | 36.66* |
| H4 | Assimilation with curriculum | 0.53 | 0.361 | 35.06* |
| H5 | Instructors' community development | 0.70 | 0.377 | 37.66* |
| H6 | Blended learning | 0.63 | 0.491 | 59.89* |

*Significant at P < 0.05.

Within this PLS methodology, the dependent variable of the research model (m-Learning adoption according to the instructors' perceptions) is considered as the response variable, and the independent variables (CSFs) are considered as predicators. The statistical results, which contain the observed values of the path coefficient R2 and the F-ratio, are illustrated in Table 3. The technical competence of instructors is observed to be significant at $P < 0.05$, with a path coefficient of 0.64, an R2 value of 0.47, and an F-ratio of 55.89. Instructors' autonomy has a path coefficient of 0.64, an R2 value of 0.43, and an F-ratio of 47.35. User-friendly design has the same direction as proposed in hypothesis H3, with a path coefficient of 0.58, an R2 value of 0.37, and an F-ratio of 36.66. Assimilation with curriculum has the same direction as proposed in hypothesis H4 with a path coefficient of 0.53, an R2 value of 0.36, and an F-ratio of 35.06. Instructors' community development has a path coefficient of 0.70, an R2 value of 0.37, and an F-ratio of 37.66. Finally, the variable of blended learning has a path coefficient of 0.63, an R2 value of 0.49, and an F-ratio of 59.89. All the corresponding p-values related to F- ratios have been found significant at $< 0.05$.

## 6.3. Testing of the Research Model

A multiple linear regression equation for our research model was presented earlier in Eq-1. In order to determine the coefficients of the equation above we ran a multiple regression analysis. In addition to giving the model coefficient, the regression also gives the direction of association. As can be seen from the model Eq-1, all the CSFs are assumed to have a positive association with user perception. The regression analysis will inform whether this is true in all cases. The results are given in Table 4 below.

Table 4.  Multiple regression analysis of the research model.

| Critical Success Factor | Coefficient term | Coefficient value | t-value |
|---|---|---|---|
| Technical competence of instructors | f1 | 0.217 | 1.62* |
| Instructors' Autonomy | f2 | 0.237 | 1.55* |
| User friendly design | f3 | 0.089 | 0.57** |
| Assimilation with curriculum | f4 | 0.120 | 0.93** |
| Instructors community development | f5 | -0.222 | -1.04** |
| Blended learning | f6 | 0.336 | 2.71* |

*Significant at $P < 0.05$. ** Insignificant at $P >= 0.05$

Table 4 shows the multiple regression calculation results. The t-value for the technical competence of instructors and m-Learning adoption according to instructor's perceptions was positive, 1.62 at $P < 0.05$, and, hence, it is significant. For the t-value between instructors' autonomy and m-Learning adoption according to instructors' perceptions was positive at 1.55 at $P < 0.05$, which indicates that it is significant. Besides, the t-value was 0.57 and the $P > 0.05$, illustrative of a statistically insignificant relationship between user-friendly design and m-Learning adoption according to instructors' perceptions. For H4, the relationship between assimilation with curriculum and m-Learning adoption according to instructors' perceptions, the t-value was 0.93 at $P > 0.05$, which means it was found to be insignificant, too; consequently, it was also rejected. Similarly, hypothesis H5 was rejected since the t-value was found to be negative (-1.04) at $P > 0.05$, which represents the relationship between instructors' community development and m-Learning adoption according to instructors' perceptions. The last hypothesis, H6, was accepted, based on the t-value of 2.71 at $P < 0.05$, which represents the relationship between blended learning and m-Learning adoption according to instructors' perceptions. Consequently, the results of the regression analysis offer interesting insights into the model. First, not all of the coefficients are positive. This means that CSFs – user friendly design, assimilation with curriculum, and instructors' community development – all have negative association with instructors' perceptions. This deviates from the expected relationship.

The final regression equation is as follows after fitting the model in equation (1):

$$m - Learning\ adoption\ according\ to\ instructor\ preceptions$$
$$= 1.170 + 0.217\ (Techincal\ compentence\ of\ instructors)$$
$$+ 0.237\ (Instructor\ autonomy) + 0.089\ (User\ firndly\ design)$$
$$+ 0.120\ (Assimilation\ with\ curriculum)$$
$$- 0.222\ (Instructors\ community\ development)$$
$$+ 0.336\ (Blended\ learning)$$

From the regression analysis, it is seen that the model accounts for 59.79% variability in the dependent variable, i.e., instructors' perceptions.

## 7. DISCUSSION OF THE RESULTS

The use of the Internet was also universal and a majority of the population accessed the Internet from their mobile devices. The instructors were also found to be technically savvy and owned other devices such as a desktop PC, laptops, and tablet PCs. This clearly shows that lack of technical awareness is not an issue in the adoption of an m-Learning platform within five Saudi Arabia universities.

In examining our results, Data analysis was started first to assess the reliability of the instrument. This was done by conducting an internal analysis and by determining the Cronbach's alpha for these multiple-items. It was found that the Cronbach's alpha in all the cases was > 0.7. This is clearly much higher than even the recently determined higher threshold of 0.6. Hence, the averages of the response could be used for determining the individual variable coefficients in the research model.

The next step involved determining whether there was a correlation between the different independent variables and the dependent variables. In the present study, both parametric and non-parametric studies were carried out. This was to remove the threats to external validity. In all of the cases, the Spearman correlation coefficient was found to be somewhat lower than the Pearson correlation coefficient, though the correlations were always >0.4. More importantly, all the hypotheses were found to be statistically significant as the p-values in each case for both parametric and non-parametric correlation analysis were found to be < 0.05. This meant that in all cases there was a reasonable correlation between the various CSFs and the instructors' perceptions based on the current data.

Once it was determined that the CSFs had statistically significant relationships with m-Learning adoption according to instructors' perceptions, the next step was to determine the regression model. It is at this point that the present study reaches a hitch. First, in the case of the variable, instructors' community development, the expected direction is negative. This means that in all these cases, the instructors believe that the CSF is inversely related to the success of m-Learning. One of the research studies in the literature review section points towards the attitude that instructors believe that mobile phones are disruptive to m-Learning, which might explain this attitude [30]. The prime objective of this work, to find out which factors have the greatest influence, was attributed to m-Learning adoption in the perspectives of the instructors; therefore, the statistical outcome confirms that the three factors are technical competence of instructors, instructors' autonomy, and blended learning, which were found to be significant in the research model testing.

## 8. LIMITATIONS OF THE STUDY

Our study relies on an empirical investigation, this means that this study has certain limitations. In general, Wohlin et al., [31] indicated that the external validity is the most common threat that limits researchers ability to generalize their experimental result to industrial practice, which is applicable in this study, too. We have ensured that specific measures have been taken to support external validity, and that includes our use of a random sampling technique that selects respondents from all departments to at least represent the general population of instructors within the university. Another limitation of this study involves its relatively small sample size. Although we sent our survey to a large number of instructors who were teaching in five universities, we only received 64 responses. Consequently, the relatively small number of responses was another potential threat to the external validity.

Furthermore, another aspect of validity concerns whether or not the study results correspond to previous findings. Our work involved the selection of six independent variables that related to the dependent variable of instructors' perspectives. While there are other key factors that influence m-Learning adoption, the scope of this study was restricted to the area of m-Learning adoption from the perspective of instructors.

However, we followed the appropriate research procedures by conducting and reporting tests in order to improve the reliability and validity of the study, and certain measures were also taken to increase the external validity.

## 9. CONCLUSIONS

In this paper, the key contribution of this work is a systematic investigation into the CSFs affecting m-Learning adoption from the perspective of instructors. As instructors are one of the vital user groups, it is important to understand the factors they consider crucial for the success of m-Learning. The results of our study showed that, according to instructors, only three out of the six factors analyzed – the technical competence of instructors, instructors' autonomy, and blended learning – were found to be statistically significant. On other hand, user-friendly design, assimilation with the curriculum, and instructors' community development, all had insignificant association with the success of m-Learning. Finally, future research work would endeavor to build a maturity model for m-Learning based on the factors that have been found to be significant.

## ACKNOWLEDGEMENTS

The first author would like to thank the ministry of higher education in Saudi Arabia for his personal fund.

## Authors

**Mr. Muasaad Alrasheedi** is a Ph.D. Candidate in the department of Electrical and Computer Engineering (Software Engineering Program) at Western University, London, Canada. Mr. Alrasheedi has a Bachelor of Science degree in Information Technology and Computing from Arab Open University, Saudi Arabia and Master of Engineering in Technology Innovation Management from Carleton University, Canada. His research interest is in the mobile learning and emerging educational technology.

**Dr. Luiz Fernando Capretz** is a Professor of Software Engineering and Assistant Dean (IT and e-Learning) at Western University, London, Canada. His research interests include software engineering, technology-enhanced learning, human factors in software engineering, and software engineering education. Dr. Capretz has a PhD in computing science from the University of Newcastle upon Tyne. He is a senior member of the IEEE, a distinguished member of the ACM, an MBTI certified practitioner, and a Professional Engineer in Ontario (Canada).

**Dr. Arif Raza** received his PhD (2011) in Software Engineering from Western university, Canada. He has authored and co-authored over 20 research articles in peer reviewed journals and conference proceedings. His current research interests include HCI, empirical studies, and usability engineering and human factors in SE.